\documentclass{Rinton-P9x6}
\usepackage{amssymb}
\newfont\fiverm{cmr5}

\begin{document}

\newcommand{\TeV}{\,{\rm TeV}}
\newcommand{\GeV}{\,{\rm GeV}}
\newcommand{\MeV}{\,{\rm MeV}}
\newcommand{\keV}{\,{\rm keV}}
\newcommand{\eV}{\,{\rm eV}}
\def\ap{\approx}
\newcommand{\bea}{\begin{eqnarray}}
\newcommand{\eea}{\end{eqnarray}}
\def\beq{\begin{equation}}
\def\eeq{\end{equation}}
\def\haf{\frac{1}{2}}
\def\lpp{\lambda''}
\def\ccg{\cal G}
\def\slash#1{#1\!\!\!\!\!\!/}
\def\u{{\cal U}}

\setcounter{page}{1}

\title{Axino as a sterile neutrino}
 
\author{Kiwoon Choi}

\address{Korea Advanced Institute of Science and Technology,
        Taejon 305-701, Korea}

\date{\today}


\maketitle

\abstracts{
We present a supersymmetric axion model in which the fermionic
superpartner of axion, i.e. the axino, corresponds to a sterile neutrino
which would accommodate the LSND data
with atmospheric and solar neutrino oscillations.}




Current data from the atmospheric and 
solar neutrino experiments are beautifully explained  
by oscillations among three active neutrino species \cite{glb}. 
Another data in favor of neutrino oscillation has been obtained 
in the LSND experiment \cite{lsnd}.  
Reconciliation of these experimental results 
requires three distinct mass-squared differences, 
implying the existence of a sterile neutrino $\nu_s$.
In the four-neutrino oscillation framework, there are two 
possible scenarios \cite{barger,giunti}: 
the $2+2$ scheme in which  two pairs of close mass eigenstates are separated
by the LSND mass gap $\sim 1$ eV
and the $3+1$ scheme in which one mass is isolated from the other
three by the LSND mass gap. 
It has been claimed  that the LSND results can  
be compatible with various short-baseline 
experiments only in the context of the 2+2 scheme.  
However, according to the new LSND results \cite{lsnd},
the allowed parameter regions are shifted to smaller mixing angle,
thereby allowing the $3+1$ scheme to be phenomenologically
viable \cite{barger,giunti}.
Although it can be realized in a rather limited parameter space,
the 3+1 scheme is attractive since the fourth (sterile) neutrino 
can be added without changing the most favorable picture that 
the atmospheric and solar neutrino data are explained
by   the predominant $\nu_\mu \to \nu_\tau$ 
and $\nu_e \to \nu_\mu, \nu_\tau$ oscillations, respectively.
In particular,  the $3+1$ scheme with the heaviest $\nu_s$ 
would be an interesting explanation of all existing neutrino data.
In this talk, we  present a supersymmetric axion model
with gauge-mediated supersymmetry 
(SUSY) breaking in which the fermionic superpartner of
axion, i.e. the axino, corresponds to a  sterile neutrino
realizing the 3+1 scheme of four neutrino oscillation
\cite{choi}. 
In this model, axino can be as light as 1 eV, and a proper 
axino-neutrino mixing is induced by
$R$-parity violating couplings which appear as a consequence
of spontaneous $U(1)_{PQ}$ breaking. 
It turns out that only the large angle MSW solution to the solar neutrino
problem is allowed in this model.

\medskip

The model under consideration contains
three sectors: the observable sector, the SUSY-breaking sector,
and the PQ sector.  The observable sector 
contains the usual quarks, leptons,
and two Higgs superfields, {\it i.e.} the superfields of
the minimal supersymmetric standard model (MSSM).
The SUSY-breaking sector contains a gauge-singlet 
Goldstino superfield $X$ and the gauge-charged messenger superfields 
$Y,Y^c$ as in the conventional gauge-mediated SUSY-breaking 
models \cite{gmsb}.
Finally the PQ sector contains gauge-singlet superfields $S_k$
($k=1,2,3$)  which break $U(1)_{PQ}$ by their vacuum expectation 
values (VEV), 
as well as gauge-charged superfields $T,T^c$
which have  the Yukawa coupling with some of $S_k$.

\medskip

The K\"ahler potential of the model can always be written as 
\beq
K=\sum_I\Phi_I^{\dagger}\Phi_I+...,
\eeq
where $\Phi_I$ denote generic chiral superfields of the model
and the ellipsis stands for (irrelevant) higher dimensional operators
which are suppressed by some powers of
$1/M_*$ where $M_*$ corresponds to the cutoff scale
of the model which is presumed to be of order 
$M_{GUT}\sim 10^{16}$ GeV.
The superpotential of the model is given by
\beq
\label{superpotential}
W=h S_3(S_1S_2-f_{PQ}^2)+\kappa S_1TT^c+
\lambda XYY^c+ W_{\rm MSSM}+W_{\rm SB}
\eeq
where $W_{\rm MSSM}$ involves the MSSM fields, and
$W_{\rm SB}$ describes SUSY breaking dynamics
enforcing $X$ develop  a SUSY breaking VEV:
$\langle\lambda X\rangle =M_X+\theta^2 F_X$.
This VEV generates soft masses of the MSSM fields,
$m_{\rm soft}\sim\alpha F_X/2\pi M_X$, as
in the conventional gauge-mediated SUSY breaking models
\cite{gmsb}.
One can easily arrange the symmetries of the model, e.g.
$U(1)_{PQ}$ and an additional discrete symmetry, 
to make that $W_{\rm MSSM}$ is given by
\bea
W_{\rm MSSM}&&=y^{(E)}_{ij}H_1L_iE^c_j+y^{(D)}_{ij}H_1Q_iD^c_j+
y^{(U)}_{ij}H_2Q_iU^c_j
+\frac{y_0}{M_*}S_1^2H_1H_2
\nonumber \\
&&+\frac{y^{\prime}_i}{M_*^2}S_1^3L_iH_2
+\frac{\gamma_{ijk}}{M_*}S_1L_iL_jE^c_k
+\frac{\gamma^{\prime}_{ijk}}{M_*}S_1L_iQ_jD^c_k+...,
\eea
where the Higgs, quark and lepton superfields are in obvious
notations
and the ellipsis stands for (irrelevant) higher dimensional
operators.

\medskip

To discuss the effective action at scales below
$f_{PQ}$, let us define the axion superfield as
$$A=(\phi+ia)+\theta\tilde{a}+\theta^2 F_A,
$$
where $a$, $\phi$ and $\tilde{a}$ are
the axion, saxion and axino, respectively.
It is then convenient
to parameterize $S_1$ and $S_2$ as 
$S_1=Se^{A/f_{PQ}}, \, S_2=Se^{-A/f_{PQ}}$.
As will be discussed later, the VEV of
$e^{\phi/f_{PQ}}=\sqrt{S_1/S_2}$ can be determined to be
of order unity by SUSY-breaking effects.
We then have  $\langle S_1\rangle
\approx \langle S_2\rangle \approx f_{PQ}$, and
then $f_{PQ}$ corresponds to the axion decay constant
which would determine most of the low energy dynamics of axion.
After integrating out the SUSY-breaking sector
as well as the heavy fields in the PQ sector,
the low energy effective action includes the
following  K\"ahler
potential and superpotential of the axion superfield $A$,
\bea
\label{effective}
K_{\rm eff}&=&f_{PQ}^2\{e^{(A+A^{\dagger})/f_{PQ}}
+e^{-(A+A^{\dagger})/f_{PQ}}\}+\Delta K_{\rm eff},
\nonumber \\
W_{\rm eff}&=&\mu_0e^{2A/f_{PQ}}H_1H_2+
\mu^{\prime}_ie^{3A/f_{PQ}}L_iH_2
\nonumber \\ 
&& +e^{A/f_{PQ}}(\lambda_{ijk}L_iL_jE^c_k
+\lambda^{\prime}_{ijk}L_iQ_jD^c_k),
\eea
where
$\Delta K_{\rm eff}$ is {\it $A$-dependent} loop corrections
involving the SUSY-breaking effects and
\bea 
\label{Frogatt}
&& \mu_0=y_0f_{PQ}^2/M_*,
\quad \mu^{\prime}_i=y^{\prime}_if_{PQ}^3/M_*^2,
\nonumber \\
&& \lambda_{ijk}=\gamma_{ijk}f_{PQ}/M_*,
\quad 
\lambda^{\prime}_{ijk}=\gamma^{\prime}_{ijk}f_{PQ}/M_*.
\eea

\medskip

The best lower bound on $f_{PQ}$ is from astrophysical arguments
implying $f_{PQ}\gtrsim 10^9$ GeV \cite{pqbound}.
To accommodate the LSND data, we need 
the axino-neutrino mixing mass of order 0.1 eV.
It turns out that this value   
is difficult to be obtained for $f_{PQ} > 10^{10}$ GeV.
We thus assume $f_{PQ}=10^9-10^{10}$  GeV
with $M_*=M_{GUT}$
for which $\mu_0$ takes an weak scale value
(with appropriate value of $y_0$)
\cite{kim}
and the $R$-parity violating couplings
$\lambda_{ijk},\lambda^{\prime}_{ijk}$ are appropriately
suppressed.
It is also easy to make the coefficient 
$\lambda^{\prime\prime}_{ijk}$ of $B$ and $R$-parity violating
operators $U^c_iD^c_jD^c_k$ to be suppressed
enough to avoid a too rapid proton decay into
light gravitino and/or axino
\cite{pdecay}.

\medskip

Low energy properties of the axion superfield
crucially depends on how the saxion component is stabilized.
One dominant contribution to the saxion effective potential
comes from $\Delta K_{\rm eff}$
which is induced mainly by
the threshold effects of $T,T^c$ having the $A$-dependent mass 
$M_T
=\kappa f_{PQ}e^{A/f_{PQ}}$.
If $M_T\lesssim M_X$, one finds
\cite{grisaru}
\beq
\label{quantumkahler}
\Delta K_{\rm eff}
\ap  
\frac{N_T}{16\pi^2}
\frac{M_TM_T^{\dagger}}{{\cal Z}_T{\cal Z}_{T^c}}
\ln\left(\frac{\Lambda^2{\cal Z}_T{\cal Z}_{T^c}}{
M_TM_T^{\dagger}}\right),
\eeq
where $N_T$ is the number of chiral superfields in $T$,
${\cal Z}_T$ is the K\"ahler metric of $T$, and $\Lambda$
is a cutoff scale which is of order $M_X$.
With ${\cal Z}_T|_{\theta^2\bar{\theta}^2}\approx
-m_{\rm soft}^2$, $\Delta K_{\rm eff}$ of (\ref{quantumkahler})
gives a negative-definite saxion potential
\beq
V^{(1)}_\phi\approx
-\frac{N_T}{16\pi^2}m_{\rm soft}^2|\kappa f_{PQ}|^2 e^{2\phi/f_{PQ}}. 
\eeq
There is another (positive-definite) potential from  the
$A$-dependent $\mu$-parameter:
\beq
V^{(2)}_\phi\approx e^{4\phi/f_{PQ}}
|\mu_0|^2(|H_1|^2+|H_2|^2).
\eeq
With $V^{(1)}_\phi+V^{(2)}_\phi$,
the saxion can be stabilized at
$\langle e^{\phi/f_{PQ}}\rangle \ap 1$
when $\kappa$ is of order $10^{-6}$.
Once $\phi$ is stabilized at $\langle e^{\phi/f_{PQ}}\rangle
\approx 1$,
the resulting  saxion and axino masses are given by
\bea
\label{axinomass}
m^2_\phi &&
\ap  \, (10 - 10^2 \, {\rm keV})^2 +
\Delta m_\phi^2,  \nonumber \\
 m_{\tilde{a}} &&
\ap \,
(10^{-4}- 10^{-2} \, {\rm eV})+\Delta m_{\tilde{a}},
\eea
where the numbers in the brackets represent the gauge-mediated contributions
for $f_{PQ}=10^9 - 10^{10}$ GeV,
$\mu_0\ap 300$ GeV and
$\langle e^{\phi/f_{PQ}}\rangle\ap 1$, and
$\Delta m_\phi$ and $\Delta m_{\tilde{a}}$
are the supergravity-mediated contributions
which are of order the gravitino mass $m_{3/2}$ 
as will be discussed in the subsequent paragraph.

\medskip

The supergravity-mediated contributions to the saxion and axino
masses can be quite model-dependent, in particular depends on
the couplings of light moduli in the underlying supergravity
model. However they are still
generically of order the gravitino mass $m_{3/2}$ \cite{lukas}. 
One model-independent supergravity-mediated contribution
is from the auxiliary component $u$ of the off-shell supergravity multiplet.
In the Weyl-compensator formulation, $u$ corresponds to
the $F$-component of the Weyl compensator superfield:
\beq
\Phi=1+\theta^2 F_{\Phi},
\eeq
where the scalar component of $\Phi$ is normalized to be unity
and the $F$-component is given by \cite{rs}
\beq
F_{\Phi}=e^{K/6}\left(m_{3/2}+\frac{F_I}{3}\frac{\partial K}{\partial \Phi_I}\right)
\eeq
where
$F_I=-e^{-K/2}K^{IJ}\partial (e^KW^{\dagger})/\partial \Phi^{\dagger}_J$
denotes the $F$-component of 
$\Phi_I$ for the inverse K\"ahler metric $K^{IJ}$ which
is determined by the K\"ahler potential $K$ of the
underlying supergravity model.
Note that $\Phi$ is defined as a dimensionless superfield,
so $F_{\Phi}$ has mass-dimension one.
The above expression shows that
$F_{\Phi}$ is generically of order $m_{3/2}$.
However it can be significantly smaller than $m_{3/2}$
in some specific models. For instance, in no-scale model
with $K=-3\ln (T+T^{\dagger}-\Phi_i\Phi_i^{\dagger})$ and
$\partial W/\partial T=0$, one easily finds
$F_{\Phi}=0$.

\medskip

The Weyl-compensator contribution to the saxion and axino masses
can be easily read off from the super-Weyl invariant supergravity action 
on superspace \cite{rs}: 
\beq
-3\int d^4\theta  \, \Phi\Phi^{\dagger}
e^{-K/3}+[\int d^2\theta \, \Phi^3 W+{\rm h.c.}]
\eeq
This gives the following couplings of $\Phi$ to
the axion superfield:
\beq
\int d^4\theta \,  \Phi\Phi^{\dagger} K_{\rm eff}
=\int d^4\theta \, \Phi\Phi^{\dagger}(A+A^{\dagger})^2+...,
\eeq
where $K_{\rm eff}$ is the effective K\"ahler potential 
in (\ref{effective}).
It is then straightforward to see that the Weyl compensator
contributions to the saxion and axino masses are 
\bea
&& \Delta m_\phi^2=
2|F_{\Phi}|^2 ={\cal O}( m^2_{3/2}),
\nonumber \\
&&
\Delta m_{\tilde{a}}
=F_{\Phi}={\cal O}(m_{3/2}).
\eea 

\medskip
In gauge-mediated SUSY breaking models \cite{gmsb}, 
the precise value of $m_{3/2}$ depends on the details of 
SUSY breaking sector. However most of models give
$m_{3/2}\gtrsim 1$ eV, implying
that $m_{\tilde{a}}$ of Eq. (\ref{axinomass}) is {\it dominated by
the supergravity contribution $\Delta m_{\tilde{a}}$.}
In this paper, we assume that $\Delta m_{\tilde{a}}\sim
1$ eV, so 
\beq \label{mss}
 m_{\tilde{a}} \approx 1 \, \,  {\rm eV}
\eeq
which would allow the axino
to be a sterile neutrino for the LSND data.
We note again that although  it is generically of order $m_{3/2}$,
$\Delta m_{\tilde{a}}$ can be significantly smaller
than $m_{3/2}$ when the supergravity K\"ahler potential takes 
a particular form, e.g. the no-scale form \cite{lukas}.

\medskip

Having defined our supersymmetric axion model,
it is rather straightforward to compute 
the $4\times 4$ axino-neutrino mass matrix:
\begin{equation} \label{Lmass}
\frac{1}{2}m_{\alpha\beta} \nu_\alpha \nu_\beta
\end{equation}
where $\alpha,\beta=s,e,\mu,\tau$ and $\nu_s\equiv \tilde{a}$
with $m_{ss}=m_{\tilde{a}}$.
The effective superpotential $W_{\rm eff}$ in (\ref{effective}) gives 
the following superpotential couplings
\beq
\int d^2\theta \,
\left[ \mu_0(1+\frac{2A}{f_{PQ}})H_1H_2
+\mu^{\prime}_i(1+\frac{3A}{f_{PQ}})L_iH_2\right].
\eeq
We will work in the field basis in which
$\mu^{\prime}_iL_iH_2$ 
$(i=e,\mu,\tau)$ in $W_{\rm eff}$ are {\it rotated away}
by an appropriate unitary rotation of $H_1$ and $L_i$.
After this unitary rotation,
the above superpotential couplings are changed to 
\beq
\int d^2\theta \, \left[\mu_0 (1+\frac{2A}{f_{PQ}})H_1H_2
+\frac{\mu_i^{\prime}A}{f_{PQ}}L_iH_2\right],
\eeq
leading to the  axino-neutrino
mass mixing 
\beq \label{msi}
m_{is}=\frac{\epsilon_i\mu_0\langle H_2\rangle}{f_{PQ}}
 \approx 0.1 \left(\frac{\epsilon_i}{10^{-5}}\right)
  \left(\frac{\mu_0}{600 \,{\rm GeV}}\right)
\left(\frac{10^{10} \, {\rm GeV}}{f_{PQ}}\right) \, {\rm eV},
\eeq
where $\epsilon_i=\mu^{\prime}_i/\mu_0$. 

\medskip

The $3\times 3$ mass matrix of active neutrinos
is induced by $R$-parity violating couplings.
At tree-level,
\beq
m_{ij}\approx \frac{g_a^2\langle \tilde{\nu}^{\dagger}_i\rangle
\langle \tilde{\nu}^{\dagger}_j\rangle}{M_a},
\eeq
where  $M_a$ denote the gaugino masses.
The sneutrino VEV's $ \langle \tilde{\nu}_i\rangle$
 are determined by
the bilinear $R$-parity violations  in the SUSY-breaking
scalar potential:
$m^2_{L_iH_1}L_iH^{\dagger}_1+B^{\prime}_iL_iH_2$.
In our model, nonzero values of $m^2_{L_iH_1}$  and $B^{\prime}_i$
at the weak scale arise through  
renormalization group evolution 
(RGE), mainly by the coupling $\lambda'_{i33}y_b$ 
where $y_b$ is the $b$-quark Yukawa coupling \cite{hwang}.
Moreover, $BH_1H_2$ arises also through RGE which
predicts a large $\tan\beta\approx 40-60$
\cite{sarid}.  
We then find \cite{hwang}
\begin{equation} \label{mij}
m_{ij}\approx 10^{-2} t^4 
 \left({\lambda'_{i33}y_b\over 10^{-6}}\right)
\left({\lambda'_{j33} y_b \over 10^{-6}}\right) 
        \, {\rm eV}  
\end{equation}
where $t=\ln(M_X/m_{\tilde{l}})/\ln(10^3)$ for
the slepton mass $m_{\tilde{l}}$.
Here we have taken
$m_{\tilde{l}} \approx 300$ GeV and $\mu_0 \approx 2 m_{\tilde{l}}$
which has been suggested to be the best parameter range for correct
electroweak symmetry breaking \cite{sarid}.

\medskip

Let us see how nicely all the neutrino masses and mixing
parameters are fitted in our framework.
The analysis of Ref. [4] leads to 
the four parameter regions, R1--R4 of Table I,
accommodating the LSND with short
baseline results.
In our model, Eqs.~(\ref{mss}) and (\ref{msi})
can easily produce the LSND mass eigenvalue
$m_4 \approx m_{ss}=m_{\tilde{a}} \sim 1$ eV
and also the LSND oscillation amplitude 
\begin{equation}
 A_{LSND}=4U_{e4}^2 U_{\mu4}^2 \approx  
  4 \left(m_{es}\over m_{ss}\right)^2 
    \left(m_{\mu s}\over m_{ss}\right)^2 
\end{equation}
as the four mixing elements 
$U_{\alpha 4}$ of the $4\times 4$ mixing matrix
$U$ are given by $U_{i4} \approx m_{is}/m_{ss} \ap 0.1$
($i=e,\mu,\tau$) 
and $U_{s4}\approx 1$.
The masses and mixing of three active neutrinos can be easily analyzed by
constructing  the {\it effective} $3\times 3$ mass matrix given by
\begin{equation} \label{meff}
 m^{\rm eff}_{ij}=m_{ij}- {m_{is}m_{js} \over m_{ss}}.
\end{equation}
Upon ignoring the small loop corrections,
this mass matrix has rank two,  and can be written as
\newcommand{\x}{{\rm x}}
\newcommand{\y}{{\rm y}}
\newcommand{\z}{{\rm z}}
\newcommand{\w}{{\rm w}}
\begin{equation} \label{ses}
 m^{\rm eff}_{ij} = m_\x \hat{\x}_i  \hat{\x}_j
    +  m_\y \hat{\y}_i  \hat{\y}_j 
\end{equation}
where $\hat{\x}_i$ and $\hat{\y}_j$ are the unit vectors 
in the direction of $m_{is}$ and $\langle \tilde{\nu}_j \rangle$,
respectively.
Remarkably,  the mass scale $m_\x\approx (m_{is}/m_{ss})^2 m_{ss} 
\sim 10^{-2}$ eV gives the 
right range of the atmospheric neutrino mass.
Eq.~(\ref{mij}) shows that $m_\y$ is also
in the range of $10^{-2}$ eV,
so $m^{\rm eff}$ would be able to provide the right 
solar neutrino mass {\it unless} $\Delta m^2_{sol} \ll 10^{-4}$ eV$^2$.
Note from Eq.~(\ref{Frogatt})
that the typical size of $\epsilon_i, \lambda_{ijk}, \lambda'_{ijk}$ 
is around $10^{-6}$  for $f_{PQ}\approx
10^{10}$ GeV and $M_*\approx 10^{16}$ GeV.

\medskip

The effective
mass matrix $m^{\rm eff}_{ij}$ gives the mass eigenvalues
\begin{equation}
 m_{2,3} = {1\over2} \left(m_\x + m_\y \pm 
             \sqrt{ (m_\x + m_\y\cos^22\xi)^2+m_\y^2\sin^22\xi} \right)  
\end{equation}
and also the $3\times 3$ mixing matrix of active
neutrinos 
\begin{equation}
U_{3\times 3}= ( \hat{\z}^T , 
       \hat{\w}^T c_\theta - \hat{\x}^T s_\theta ,
       \hat{\w}^T s_\theta + \hat{\x}^T c_\theta)  \,.
\end{equation}
where $\hat{\z}\equiv \hat{\x}\times\hat{\y}/|\hat{\x}\times\hat{\y}|$,
$\hat{\w}\equiv \hat{\x}\times \hat{\z}/|\hat{\x}\times \hat{\z}|$,
$c_\xi\equiv \cos\xi = \hat{\x}\cdot\hat{\y}$
and $ \tan2\theta=  m_\y \sin2\xi/( m_\x+m_\y\cos2\xi) $.
The Super-Kamiokande data \cite{sk-atm} combined with the CHOOZ 
result \cite{chooz} imply that $U_{\mu3}^2\approx U_{\tau3}^2
\approx 1/2$ and $U_{e3}^2\ll1$.  
The solutions to the solar neutrino problem can have either a 
large mixing angle (LA): 
$U_{e1}^2\approx U_{e2}^2 \approx 1/2,$
or a small mixing angle (SA): 
$U_{e1}^2 \approx 1.$ 
This specify the first column $\hat{\z}^T$ of $U|_{3\times 3}$ as
$$
({\rm LA}): \,
\hat{\z} \approx (1/\sqrt{2}, -1/2, 1/2),
\quad 
({\rm SA}): \,
\hat{\z}\approx (1, 0, 0)
$$ 
up to sign ambiguities.  
Since $\hat{\x}\cdot\hat{\z}=0$,
the pattern $\hat{\z}\approx(1,0,0)$ implies $\hat{\x}_e \approx 0$.
This leads to a too small $U_{e4}\approx m_{es}/m_{ss}\lesssim
10^{-2}$, so the SA solution is {\it not allowed} within our model.
Among various LA solutions to the
solar neutrino problem, {\it only} the large-angle MSW solution
with $\Delta m^2_{sol} \sim 10^{-4}\,{\rm eV}^2$ 
can be naturally fitted  since
$m_\x \approx m_\y \sim 10^{-2}$ eV in our scheme.
It is remarkable  that 
$f_{PQ}\approx 10^{10}$ GeV and $M_*\approx M_{GUT}$
lead to the right size of $R$-parity violation
yielding the desired values of $m_{is}$ and $m_{ij}$
also for the atmospheric and solar neutrino masses.

\medskip
To see the feasibility of our whole scheme,
we scanned our parameter space which consists of
$m_{ss}, m_{is}, \lambda^{\prime}_{i33}y_b$
to reproduce the allowed LSND islands R1--R4 of Table I
together with the atmospheric and solar neutrino parameters \cite{choi}.
For R1 and R4, we could find some limited parameter spaces
which produce the corresponding oscillation parameters,
however  they need a strong alignment between 
$\hat{\x}$ and $\hat{\y}$ and 
also a large cancellation between $m_{\x}$ and $m_{\y}$.
On the other hand,  R2 and R3
do not require any severe fine tuning of parameters,
so a sizable range of the parameter space can fit
the whole oscillation data.

\medskip

To conclude, we have shown that the 3+1 scheme of
four-neutrino
oscillation can be nicely obtained in supersymmetric 
axion model with gauge-mediated 
supersymmetry breaking.
In this model, axino plays
the role of sterile neutrino
by having a mass $\sim 1$ eV and also
a proper axino-neutrino mixing 
induced by $R$-parity violating couplings.
One interesting feature  of the model
is that  only the large angle MSW solution to
the solar neutrino problem is allowed in this model.

\bigskip

{\bf Acknowledgement}:
This work is  supported by BK21 project of the
Ministry of Education, KOSEF through the CHEP of KNU,
KRF Grant No.~2000-015-DP0080, and KOSEF Grant No.
~2000-1-11100-001-1.

%


%
\begin{table}
\centering
\label{lsnddata}
\begin{tabular}{|c|c|c|c|}
\hline
& $|\Delta m_{41}^2|({\rm eV}^2)$ & $|U_{e4}|$ & $|U_{\mu 4}|$ \\ \hline
R1 &  0.21-0.28 & ~0.077-0.1~ & ~0.56-0.74~ \\ \hline
R2 &  0.88-1.1 & ~0.11-0.13~ & ~0.15-0.2~ \\ \hline
R3 & 1.5-2.1 & ~0.11-0.16~ & ~0.09-0.14~ \\ \hline
R4 & 5.5-7.3 & ~0.13-0.16~ & ~0.12-0.16~ \\
\hline
\end{tabular}
\vspace{0.5cm}
\caption{Allowed regions for the LSND oscillation.}
\end{table}

\end{document}